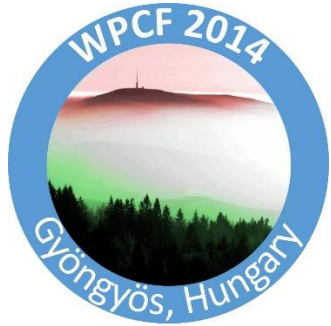

# Quark Wars

a particle physics outreach game in the age of Star Wars

T. Csörgő[1,2] and T. Novák[1]

[1]EKE KRC, H - 3200, Gyöngyös, Mátrai út 36, Hungary

[2]Wigner RCP, H-1121 Budapest XII,

Konkoly-Thege Miklós út 29-33, Hungary

December 16, 2016

**Abstract:**

Quark Wars is an all-new, adventure style game. We recommended playing it outdoors. Quark Wars is modeled upon the outdoor game called Hungarian Number War, with notable influence from Star Wars, the American epic space saga. The players form two opposing teams. Both teams elect their own leader. The team members and their leaders wear particle war bonnets on their foreheads. These headdresses consist of three or four cards indicating combinations of elementary particles. The two teams compete by identifying (reading out loudly) the elementary particle cards on the foreheads of their adversaries. Players are allowed to use the terrain to cover their particle identity on their foreheads and may try to hide, run or band together in a group to win. Quark Wars was tested at a Summer Camp of Berze Science Club in Hungary. Students loved playing Quark Wars, as this game resulted in lots of hilarity and action. In addition, Quark Wars also solidified particle terminology and made the concept of particle identification and discovery more tangible to secondary/middle school students.

# 1. Long time ago in a small village far, far away…

*It is a period of particle war. Rebel student groups, striking from their hidden base, have won their first victory against the evil particle empire.*

*During the battle, student spies managed to steal secret plans to the Empire's ultimate weapon, the Quark Star, an armored particle accelerator with enough power to destroy an entire Planet.*

*Pursued by the Empire's sinister agents, Princess Lepton races home, custodian of the stolen plans that can save her people and restore asymptotic freedom to their quarks and to the Early Universe…*
*/Opening Crawl, with apologies to George Lucas [1]/*

It was a roguish and hilarious day with a new game for the students in the Summer Camp of the Berze Secondary School in a small Hungarian village called Visznek. They had just learned about elementary particles and soon their survival depended on their ability to identify them. They were assigned to a difficult mission by their scientist patrons, namely to participate in the test of a new outdoor game called Quark Wars. This required that students develop not only some basic knowledge of elementary particles and focus their mind on some fundamental particle physics, but also tested their endurance, creativity, sneaking skills and fortitude.

The students were already introduced to some of the basic concepts and terminology of elementary particle physics, thanks to their familiarity with certain type of Quark Matter Card Games [2]. In these games, elementary particles like quarks and leptons are represented by particle cards. Csaba Török, a 17-year-old Hungarian secondary school student conceived the idea of a deck of elementary particles, and he created ANTI, the first of the Quark Matter Card Games [3], on the New Year's Eve of 2008/2009. Csaba was a member of the very same Berze Science Club, where the Quark Wars was first tested. Csaba had heard several outreach talks about particle physics in this Science Club before had the idea of representing quarks and leptons of the Standard Model on the faces of a deck of cards. His first game, ANTI was subsequently refined and developed to the so called Quark Matter Card Game [4] with the primary goal of entertainment and secondary goal as a science outreach tool. In this development, Csaba teamed with Judit Csörgő, another 17-year-old member of the Berze Science Club, who realized that these games could be taught even to pre-school 5-years-olds. Tamás Csörgő, Judit's father (and one of the co-authors of this article) also joined the team as a mentor and manager. He is a research physicist who works on experimental as well as theoretical high energy physics problems related to the RHIC accelerator at Brookhaven National Laboratory, USA and the LHC accelerator at CERN. Tamás also acts as the scientist patron of the Berze Science Club.

The Quark Matter Card Game of J. Csörgő, Cs. Török and T. Csörgő [2] provides several opportunities to have fun with elementary particles and anti-particles, in order to model the Big Bang, the formation of the Early Universe in the first few microseconds after its creation, or to play Quark Matter to model and popularize the time evolution of heavy ion collisions at RHIC and at LHC [2], or to memorize quarks [5] or to search for your own Higgs boson [6]. Each of these games can be started on the level of laypersons, in an entertaining and delightful manner, in a way that triggers a learning spiral as the players become more and more enthusiastic and sophisticated participants. The games are designed to scale well with the knowledge level of the participants, as we shall demonstrate here as well on the new game called Quark Wars.

## 2 Preparations and inventory

In order to be able to play Quark Wars, the students were first of all reminded about some fundamental concepts of elementary particle physics, as realized in the form of the Quark Matter Card Game. They were reminded that in these games two main groups of particles are included: quarks and leptons. Quarks are represented with cards colored red, green or blue, indicating that they participate in the strong interactions. According to Quantum Chromo Dynamics, the theory of the strong interactions, quarks and their anti-particles, called anti-quarks, have certain symmetry properties similar to that of the optical colors. Leptons are charged or neutral particles who do not participate in the strong (or color) interactions, so leptons are represented by black and white cards in the Quark Matter Card Games.

As each particle has an anti-particle, the question arises, how to represent (model) anti-particles on the faces of Quark Matter Cards. We emphasized that for leptons with electric charge, positive and negative signs indicate the electric charge of particle / anti-particle pairs, for example an electron is denoted by " $e^-$ ", so its anti-particle, the positron is denoted by " $e^+$ ". Obviously, for the quarks with red, green, blue "color charges", and the anti-quarks should be modeled by the corresponding anti-colors.

But what are the anti-red, anti-green and anti-blue colors? In the model applied in the Quark Matter Card Games, anti-color is what makes a given color neutral i.e. white. This way, the anti-red is defined as a green/blue combination, while anti-green is blue/red and anti-blue is red/green. An important rule in the Quark Wars game is that quarks and anti-quarks can appear only in a color neutral or white combination of colors: a red, a green and a blue quarks may form a colorless, white particle called a baryon, a red and anti-red pair of quarks and anti-quarks (or a blue-anti-blue quark – anti-quark pair or a green-anti-green quark – anti-quark pair) forms a meson. The left panel of Figure 1 shows the students participating in this training section. The main idea of the game is indicated on the right panel of Figure 1, where one of their mentors, T. Novák is shown wearing a color neutral combination of quarks on his forehead as a kind of Particle War Bonnet.

Simple tools that are available everywhere can be used to transform the Quark Matter Cards to Particle War Bonnets: a page of a plastic name card holder, elastic rubber bands and cards representing elementary particle will do the job. A valid combination of quarks and anti-quarks or a pair of leptons and anti-leptons are inserted in the same page of the plastic name card holder, the page is cut out and fixed to the forehead of the players as indicated on Figure 2.

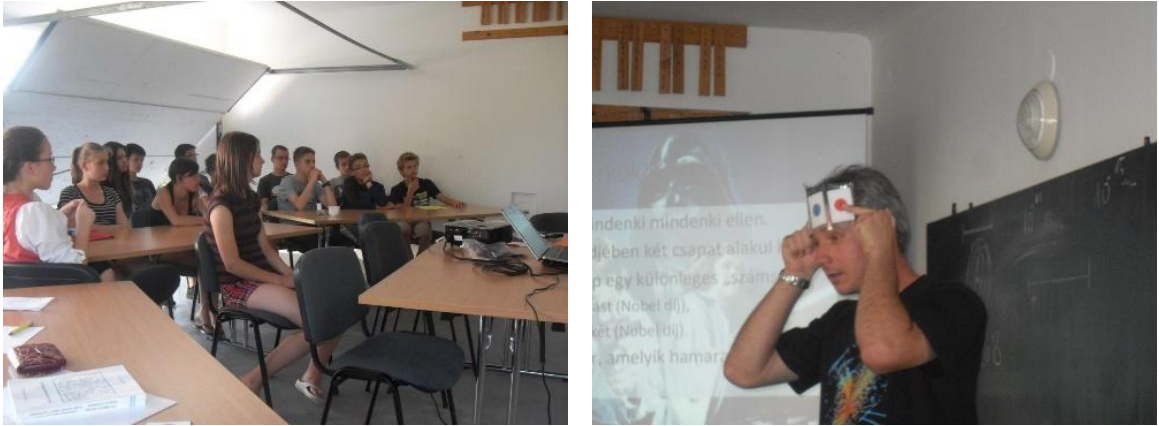

**Figure 1.** Science Club students being introduced to Quark Wars at the Summer Camp (left panel). One of the authors before the screen, wearing Quark Matter Cards as a kind of Particle War Bonnet on his forehead to demonstrate the main idea of this game.

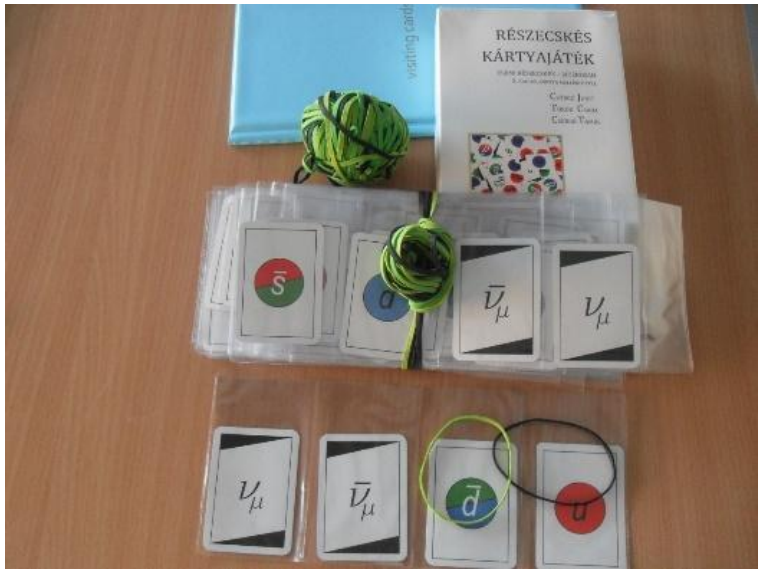

**Figure 2.** Only a few things are needed to play Quark Wars: elastic rubber bands and a name card holder to keep the cards in position, and a deck of Quark Matter Card Game. The deck is included, for example in the Appendix of the booklet on Quark Matter Card game, shown in Hungarian here. This booklet is available and downloadable in English as well [2], suitable for printing, so that anybody with a color printer can prepare his or her own particle war bonnets for Quark Wars. The two competing teams are recommended to indicate their team, e.g., with the color code of the rubber band (here green or black) that is placed on their foreheads to keep the cards in place.

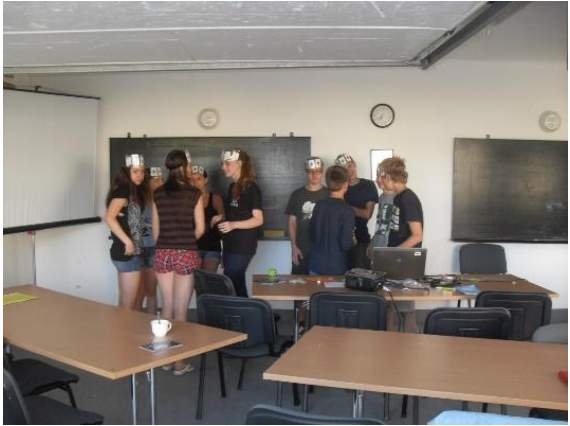 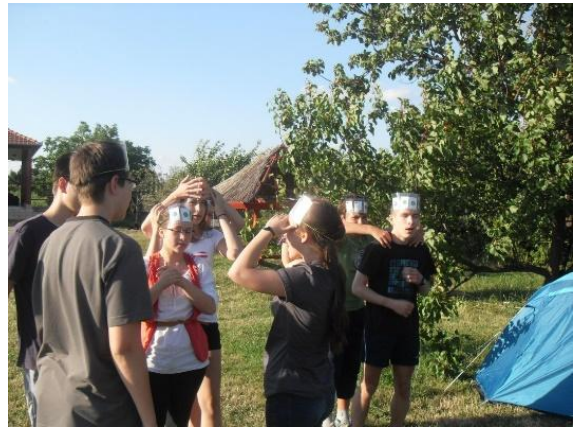

**Figure 3.** Two teams are formed, they have a brief (2-3 min) council on how they plan to win the game. Then the players put on their particle war bonnets and hide their team leaders so that members of the other team could not easily find and identify them.

The players form two teams and receive their quark matter war bonnets. The two teams separate and each have a council: team members come together to consult and to make decisions, to develop their plans and strategies on how to play and win the game. Then players put on their Quark Matter Card Game War Bonnets, but in such a way that members of the opposing team could not spy on them as seen on Figure 3, and start the actual Quark Wars.

## 3 Quark Wars – the goal and the course of the game

The **number of players** is, in principle, arbitrary, but in practice, it is best if both teams have 5 members or more, limited by available number of children in a classroom or in a summer camp.

The **object of the game** is to win, which can be either by identifying all players of the opposing team by leading out loudly the combination of elementary particles on their particle war bonnets, or, by finding and identifying (reading out loudly) the particle identity of the leader of the opposing team.

Due to this mission, during the **course of the game** it is **recommended** that the team leaders wear special particle war bonnets that correspond to Nobel prize winning elementary particles such as the $H^0$ Higgs boson [7,8] or the $\Omega^-$ baryon. When the fundamental theoretical predictions on the existence of these elementary particles were confirmed by experimental observations, Nobel prizes in physics were awarded to Peter W. Higgs and Francois Englert in 2013 [9], and to Murray Gell-Mann in 1969 [10], respectively. So identifying either a $H^0$ or an $\Omega^-$ during the course of Quark Wars identifies one of the team leaders and abruptly ends the Quark Wars game.

This goal also makes the game an interesting strategic game: when one of the teams has only a few members left, they can still try to break into the area of the enemy and to find a weakness in the defense of their opponents. Identifying the particle signs weared by the leader of the opposing team may result in winning the game, similarly how a small squadron of rebels may win the battle against a big army of imperial troops in the epic space opera Star Wars by attacking the Death Star on its weakest point. Due to this reason one of the team leaders can also be named informally as Luke Skywalker and the other as Darth Vader, however, both must wear their particle war bonnet during Quark Wars and can be identified on their particle name only. Identified players become inactivated, they cannot continue their participation in particle hunting anymore, they become silent observers during the subsequent course of the game.

Note that light-sabers and other toy weapons are not permitted in the Quark Wars style game, not only for safety reasons, but also because we followed the rules of the so called Hungarian Number Wars [11], a game that was familiar to the participants of the Summer Camp of the Berze Science Club in Hungary. In this kind of games, the players are not allowed to cover their war bonnets on their foreheads with their hands. It is also not allowed to "worship the Earth" by lying down and pressing foreheads to the ground: this usually leads to identification by a group from the opposing team, who force the Earth-worshipper to turn up and identify him or her, but this strategy is not only unfair but also makes the game less interesting. The participants are allowed to cover themselves by pressing their foreheads to objects on the terrain such as trees, columns, buildings, and they are allowed to band together in groups of 2-4 to try to win, as indicated in Figure 4.

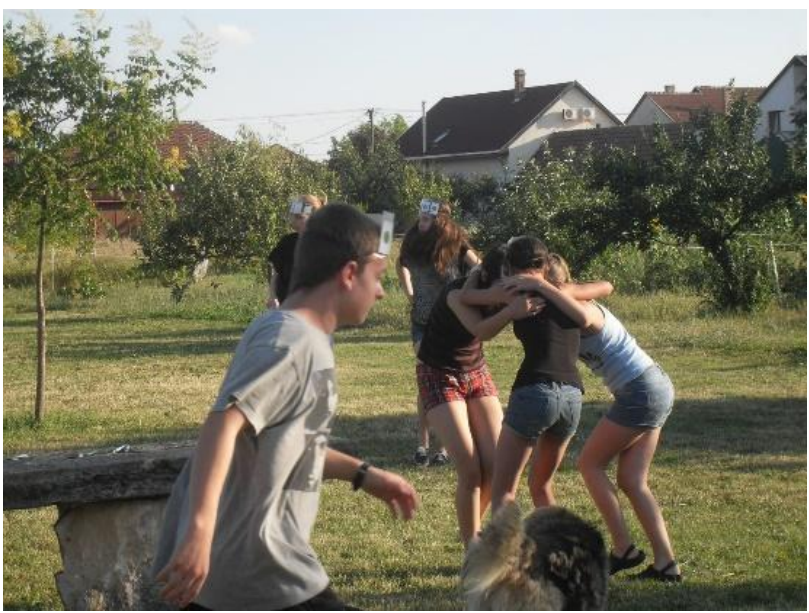

**Figure 4.** Players hiding their particle identities on their foreheads. One of the strategies is rushing: running forward while shaking their heads. Another, safer but slower strategy is to band in a group of 2, 3 or 4, pressing foreheads together. This method allows for a slower progress, but it also minimizes the possibility of being identified.

Much of the fun of the Quark Wars comes from the strategies that the players apply to hide their particle identities on their foreheads. One of the possible good strategies is rushing (running forward and shaking head simultaneously) as indicated on Figure 4. Rushing is designed so that reading out the card combination be difficult for the opponents. This allows fast progress but it is risky and physically challenging, needs practice for a good performance. Another, safer strategy is to band in a group of 2, 3 or 4, pressing foreheads together. This method allows for slow progress, and it minimizes the possibility of being identified. The drawback of banding is that it also makes it very difficult to identify anybody from among the opposing team, as the members of the band can hardly see from one another. The best seems to move so that the foreheads are covered with the objects on the terrain as much as possible, to move slowly if covered and to rush through the open areas to reach the terrain of the opponent team.

While the battle is ongoing, our heroes, the leaders of the teams may be just hiding in a safe place, to minimize the risk that they are identified and the other members of the teams try to protect their hiding places, chosen to be difficult to guess or figure out. As an example, during a test Quark War game, one of the team leaders, with particle war bonnet indicating an $\Omega^-$ particle was hiding behind the pull-down screen of the instruction room, as shown in the left panel of Figure 5. The other team leader, with a war bonnet representing a Higgs boson, was hiding in the middle of a haystack as indicated on the right photo of Figure 5. Eventually, all the team members of the Emperor's "Higgs boson" team were identified and hence removed from active participation by the Rebel team, the members of the team of $\Omega^-$. Even in this period, it was quite a challenge for the surviving members of the Rebel Alliance or team $\Omega^-$ to find the hiding location of commander Darth Vader alias Higgs boson. Probably this was the first time in the history of particle physics outreach, that a symbolic Higgs boson was found and finally identified in a real hay-stack.

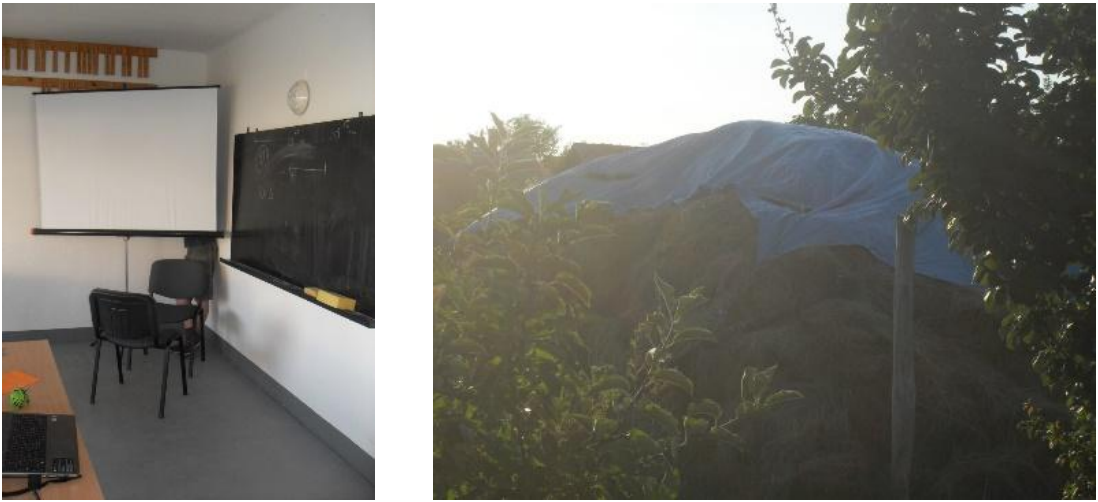

**Figure 5.** One of the team leaders, wearing a particle war bonnet of $\Omega^-$, is hiding behind the pull-down screen, while the other team leader, with particle war bonnet standing for a Higgs boson ($H^0$) is hiding in a hay-stack.

It is customary that each player keeps a note of his or her own particle identity at hand, so that they could easily cross-check if their identification attempts were successful or not.

Valid combinations on the particle war bonnets include three quark combinations representing baryons (a triplet, combination of a <span style="color:red">red</span>, a <span style="color:blue">blue</span>, and a <span style="color:green">green</span> quark) anti-baryons (a combination of three anti-quarks, with <span style="color:green">anti</span>-<span style="color:red">red</span>, <span style="color:red">anti</span>-<span style="color:blue">blue</span> and <span style="color:blue">anti</span>-<span style="color:green">green</span> colors) as well as four card combinations. These four cards consist of two valid pairs, each of the pairs is a lepton-antilepton pair or a meson consisting of a quark and an anti-quark, where the color of the quark matches the anti-color of the corresponding anti-quark.

The other rules of Quark Wars follow in general the rules of **Hungarian Number War** game, which seems to be well known only in Hungary. For a reasonable English language summary of the rules of this fun outdoor game see ref. [11].

**The Quark Wars game can be played on various knowledge levels**, so it is a kind of well scalable particle physics outreach game. In this respect, Quark Wars is similar to the Quark Matter [4], Memory of Quark Matter [5], or Find Your Own Higgs Boson [6] card games.

On a **layperson's or beginner** level, the players can read out only the quark or lepton cards to identify their opponents, for example: red u, green u, blue d! Or red u, anti-red anti-u and an electron-positron pair! They should, however, remember the combinations that identify the team leaders. It is recommended that the war bonnets of the opponent teams be fixed with rubber bands of different colors, e.g. black rubbers for one of the teams and green rubbers for the other team, as indicated on Figure 2.

On an **intermediate level,** the players should be instructed that the $\Omega^-$ baryon is a combination of a red s, green s and a blue s quark. The Higgs boson is identified through its leptonic decays, e.g., as two charged lepton-antilepton pairs. The possible Higgs decays that can be used for this purpose are given in Table 1 of [6], that we recapitulate here for completeness:

| $H^0$ decay mode | Final state particles/cards | |
|---|---|---|
| $H^0 \to \gamma,\gamma$ or $Z^0 Z^0 \to$ | $e^+ e^-$ | $e^+ e^-$ |
| $\to$ | $e^+ e^-$ | $\mu^+ \mu^-$ |
| $\to$ | $\mu^+ \mu^-$ | $\mu^+ \mu^-$ |
| $H^0 \to W^+ W^- \to$ | $e^+ \nu_e$ | $e^- \bar{\nu}_e$ |
| $\to$ | $e^+ \nu_e$ | $\mu^- \bar{\nu}_\mu$ |
| $\to$ | $\mu^+ \nu_\mu$ | $e^- \bar{\nu}_e$ |
| $\to$ | $\mu^+ \nu_\mu$ | $\mu^- \bar{\nu}_\mu$ |

**Table 1.** Possible Higgs decays as represented in the Quark Wars outreach game. On the beginner/layperson level, only one of this decays is used to identify one of the team leaders. After gaining more experience with the game, any of the above four particle card combinations can be used to indicate/symbolize one of the team leaders, marking the leader with a Higgs boson decay.

On an **advanced level,** combinations like red u, green u, blue d are not sufficient and acceptable to identify particles consisting of quarks (and/or leptons). Instead, the players should know that a color-neutral (white) combination of uud quarks stands for a proton, udd stands for a neutron, (u,anti-d) stands for a positive pion etc. These intermediate and more advanced levels can be practiced only with a supervisor or referee, who has some basic knowledge of hadrons and other elementary particles, and who can step in to decide if the particle identification was correct or not. Appendix A (Naming the hadrons) of ref. [4] can be used to find the tables that describe the quark contents of various hadrons, as represented by the cards of Quark Matter Card Game. Tables in Appendix A of ref. [4] form a kind of interface from particle physics outreach to the more substantiated e.g. quantum mechanical description of the quark contents of hadrons.

These tables are not necessary to play Quark Wars on a beginner or intermediate level, but are useful to play quark wars on an advanced level. Teachers and promoters of Quark Wars should keep in mind that students prefer first of all to play and have fun, so we recommend to start the game on a layperson's level, with as many ordinary students as possible.

The players and in particular their supervisors are also **reminded to check any potential pitfalls and dangers on the terrain**. We recommend to take the children outside to play. It is a wonderful way to spend the day [12]. If no safe outdoor terrain can be secured, this game may also be played or tested indoors, like in a gymnasium or a sports hall of a school. This Quark Wars game is apparently well suited also for a future adaptation as a computer game, but most of the fun of it seems to come from real time interactions among real players, when not only the intellectual but also the physical abilities like endurance, creativity, sneaking skills and fortitude of the participants are tested and also trained during the course of Quark Wars.

## 4 Summary

A new particle physics outreach game called Quark Wars has been developed and tested successfully in Hungary. About two dozens of secondary/middle school students (both boys and girls) participated in this test. Quark Wars as a particle physics outreach game brought together ordinary people like secondary and middle school students with teachers and scientists to do extraordinary things, and in doing so, they became part of something greater than themselves. Quark Wars, this new particle physics outreach game, not only resulted in lots of hilarity, hiding, running and yelling outdoors, but participating students also learned a good deal about particle terminology and gained a first-hand experience on how difficult it might be to locate and identify a hiding Higgs boson, and how big a pleasure it is to find and identify it eventually.

## Acknowledgments

Thanks and credits are due to Ágota Lang, a physics teacher at the Széchenyi Secondary School in Sopron, Hungary: she had the first idea to create a particle physics outreach game based on the rules of the Hungarian Number Wars game. The authors thank Ildikó Pálinkás and Judit Pető for their careful reading of this manuscript and thank the sponsors of the WPCF 2014 conference in Gyöngyös, Hungary. The presentation of Star Wars at WPCF 2014 was partially supported by the he OTKA NK 101438 grant (Hungary).

**References:**

[1] Star Wars Opening Crawl, http://www.starwars.com/video/star-wars-episode-iv-a-new-hope-opening-crawl

[2] J. Csörgő, Cs. Török, T. Csörgő, *Quark Matter Card Game - Elementary Particles on Your Own* (2nd English Edition, ISBN 978-963-89242-0-9, 2011).

[3] Introduction to ANTI, a Quark Matter speed card game for two persons (in Hungarian): https://www.youtube.com/watch?v=tEpdTcwZ3xw

[4] Trailer for the Quark Matter Card Game: https://youtu.be/Sn56IEC9VSE

[5] J. Csörgő, Cs. Török, T. Csörgő: *Memory of Quark Matter,* arXiv:1303.2798 [physics.pop-ph] (2013)

[6] T. Csörgő: *Higgs boson – On Your Own*, arXiv:1303.2732 [physics.pop-ph], in Proc. EDS Blois 2013, Saariselka, Finland, September 2013

[7] ATLAS Collaboration: „*Observation of a new particle in the search for the Standard Model Higgs boson with the ATLAS detector at LHC*", Physics Letters B 716 (2012) 1-29

[8] CMS Collaboration: „*Observation of a new boson at a mass of 125 GeV with the CMS experiment at LHC*", Physics Letters B 716 (2012) 30-61

[9] The Nobel Prize in Physics 2013 was awarded jointly to François Englert and Peter W. Higgs "*for the theoretical discovery of a mechanism that contributes to our understanding of the origin of mass of subatomic particles, and which recently was confirmed through the discovery of the predicted fundamental particle, by the ATLAS and CMS experiments at CERN's Large Hadron Collider*", http://www.nobelprize.org/nobel_prizes/physics/laureates/2013/

[10] The Nobel Prize in Physics 1969 was awarded to Murray Gell-Mann *"for his contributions and discoveries concerning the classification of elementary particles and their interactions".* http://www.nobelprize.org/nobel_prizes/physics/laureates/1969/

[11] http://hungarykum.blogspot.hu/2015/10/mopdog-monday-capture-hungarian-flag.html

[12] K. Alfano: *The benefits of outdoor play.* http://www.fisher-price.com/en_US/parenting-articles/outdoor-play/the-benefits-of-outdoor-play .